\documentclass[amsmath,amssymb,preprint,showkeys]{revtex4}
\usepackage{epsfig}
\usepackage{graphicx}

\begin{document}

\title{Signal photon flux generated by high-frequency relic gravitational waves}

\author{Xin Li $^{1,2}$}
\email{lixin1981@cqu.edu.cn}
\author{Sai Wang $^2$}
\email{wangsai@itp.ac.cn}
\author{Hao Wen $^{1,2}$}
\email{wenhao@cqu.edu.cn}
\affiliation{$^1$Department of Physics, Chongqing University, Chongqing 401331, China\\
$^2$State Key Laboratory Theoretical Physics, Institute of Theoretical Physics, Chinese
Academy of Sciences, Beijing 100190, China}

\begin{abstract}
The power spectrum of primordial tensor perturbations $\mathcal{P}_t$ increases rapidly in high frequency region if the spectral index $n_t>0$. It is shown that the amplitude of relic gravitational wave $h_t$($5\times10^9$Hz) varies from $10^{-36}$ to $10^{-25}$ while $n_t$ varies from $-6.25\times 10^{-3}$ to $0.87$. High frequency gravitational waves detector that is proposed by F.-Y. Li detects gravitational waves through observing the perturbed photon flux that is generated by interaction between the relic gravitational waves and electromagnetic system. It is shown that the perturbative photon flux $N_x^1$($5\times10^9$Hz) varies from $1.40\times10^{-4}\rm s^{-1}$ to $2.85\times10^{7}\rm s^{-1}$ while $n_t$ varies from $-6.25\times 10^{-3}$ to $0.87$. Correspondingly, the ratio of the transverse perturbative photon flux $N_x^1$ to the background photon flux varies from $10^{-28}$ to $10^{-16}$.
\end{abstract}

\maketitle
\section{Introduction}
Gravitational waves (GW), as a prediction of general relativity, have not been detected directly. It is very interesting and important to search the GW both on experiments and theories. GW carry away energy. Thus, one of most important theoretical predictions of GW is that it causes the orbital decay of binary system. This is first observed in PSR 1913+16 \cite{Taylor}. GW could cause fluctuations of space. Various detectors that based on this property of GW are constructed or proposed to search the GW in different frequency bands. It includes pulsar timing arrays ($10^{-9}-10^{-7}$Hz)\cite{Hobbs}, space-based interferometers such as eLISA ($10^{-7}-10^0$Hz) \cite{LISA}, ground-based interferometers such as LIGO ($10-10^4$Hz)\cite{LIGO}.

The standard cosmological model, i.e. the $\Lambda$CDM model \cite{Sahni,Padmanabhan} is consistent with several precise astronomical observations that involve Wilkinson Microwave Anisotropy Probe (WMAP) \cite{Komatsu}, Planck satellite \cite{Planck1}, Supernovae Cosmology Project \cite{Suzuki} and etc. However, the relic gravitational waves (RGW) that were produced in the inflationary stage \cite{Riotto} of the universe have not been detected. The RGW that could generate B-mode polarizations of cosmic microwave background radiation (CMB) have very low frequency ($10^{-18}-10^{-15}$Hz). The recent joint analysis of BICEP2/Keck Array and Planck Data (BKP) \cite{BKP} gives severe constraints on B-mode polarizations of CMB. It gives an upper limit for tensor-to-scalar ratio $r_t$, i.e. $r_t<0.12$ at 95\% confidence level at the pivot scale $0.01 \rm Mpc^{-1}$. The inflation model tells us that the RGW have a broad range spreading spectrum, from $10^{-18}$Hz to $10^{10}$Hz, where the lower and upper bounds of frequency correspond to the current Hubble radius ($1/H_0$) and the vacuum energy scale during inflation ($10^{16}$GeV), respectively \cite{ZhangY}. Thus, it is expected that GW detectors---LIGO and eLISA could catch some information of RGW. At present, LIGO S5 \cite{LIGO S5} give the most severe constraint on energy density of RGW, i.e., $\Omega_{GW}<6.9\times10^{-6}$ around $100$Hz.

The power spectrum of primordial tensor perturbations that is associated with RGW can be parameterized as \cite{Planck XX}
\begin{equation}\label{tensor spectrum}
\mathcal{P}_t(k)=r_tA_s\left(\frac{k}{k_p}\right)^{n_t},
\end{equation}
where $A_s$ is the amplitude of power spectrum of primordial scalar perturbations, $n_t$ is the tensor spectral index and $k_p$ is the pivot scale. In the canonical single-field slow-roll inflation models, $n_t$ satisfies the consistency relation, i.e., $r_t\simeq-8n_t$ \cite{Riotto}. It means $n_t<0$. However, certain inflation models or its alternative models, such as inflation with axion potential \cite{Mukohyama} and the ekpyrotic model \cite{Khoury}, predict a blue tensor spectra, namely, $n_t>0$.
Thus, the tensor spectral index $n_t$ can be used to distinguish different inflation models and alternatives. In Ref.~\cite{Huang}, by combining the data of BKP and LIGO, it yields a constraint on $n_t$, namely, $n_t=-0.76^{+1.37}_{-0.52}$ at the 68\% confidence level.

One can find from the formula (\ref{tensor spectrum}) that the power spectrum of primordial tensor perturbations $\mathcal{P}_t$ varies rapidly in high frequency region. Thus, high frequency GW detector could give more severe constraint on tensor spectral index $n_t$ if $n_t>0$. F.-Y. Li \cite{FYLi} has designed a GW detector that is focused on detection of GW at high frequency GW spectra ($10^9-10^{14}$Hz). The recent researches on F.-Y. Li's high frequency GW detector can be found in Ref. \cite{FYLi1}. In this letter, we will use the data of Planck satellite \cite{Planck XIII} to show the amplitude of RGW in high frequency region ($10^8-10^{10}$Hz). Then, we will use the amplitude of RGW to obtain the experimental signals in F.-Y. Li's GW detector.

\section{Relic gravitational waves with high frequency}
The energy density of RGW is given as \cite{Turner,Zhao}
\begin{equation}\label{density RGW}
\Omega_{GW}(k)=\frac{\mathcal{P}_t(k)}{12H_0^2}\left(\dot{\mathcal{T}}(\eta_0,k)\right)^2,
\end{equation}
where $\eta_0=1.41\times10^4$Mpc is the conformal time at present epoch, $H_0$ is the Hubble constant and the dot denotes the derivative with respect to cosmic time $t$. Here, the $\mathcal{T}(\eta_0,k)$ in formula (\ref{density RGW}) that represents tensor transfer function can be approximately described as \cite{Turner,Zhao,Kuroyanagi,Watanabe}
\begin{equation}\label{transfer f}
\mathcal{T}=\frac{3j_1(k\eta_0)\Omega_{m0}}{k\eta_0}\sqrt{1+1.36\left(\frac{k}{k_{eq}}\right)+2.50\left(\frac{k}{k_{eq}}\right)^2},
\end{equation}
where $\Omega_{m0}$ denotes the matter density parameter at present epoch, $k_{eq}=0.073\Omega_{m0}h^2 \rm Mpc^{-1}$ corresponds to the Hubble radius at matter-radiation equality and $h$ denotes the reduced Hubble constant. We can obtain the time derivative of tensor transfer function from equation (\ref{transfer f}). It is given as
\begin{equation}\label{transfer ft}
\dot{\mathcal{T}}=\frac{-3j_2(k\eta_0)\Omega_{m0}}{\eta_0}\sqrt{1+1.36\left(\frac{k}{k_{eq}}\right)+2.50\left(\frac{k}{k_{eq}}\right)^2}.
\end{equation}
By making use of the tensor transfer function (\ref{transfer f}), we obtain the amplitude of RGW at present
\begin{equation}
h_t(\eta_0,k)=\sqrt{\mathcal{P}_t(k)}\mathcal{T}(\eta_0,k).
\end{equation}

\begin{figure}
\includegraphics[scale=1]{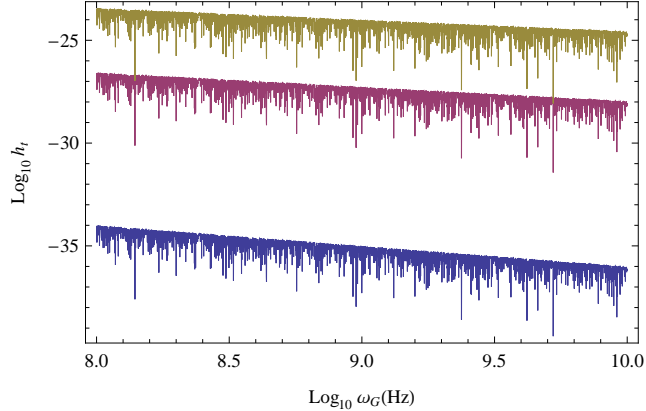}
\caption{The amplitude of Relic gravitational waves for various $n_t=-r_t/8,0.61,0.87$ that correspond to the blue curve, the purple curve and the brown curve, respectively. }
\label{fig:amplitude}
\end{figure}

\begin{figure}
\includegraphics[scale=1]{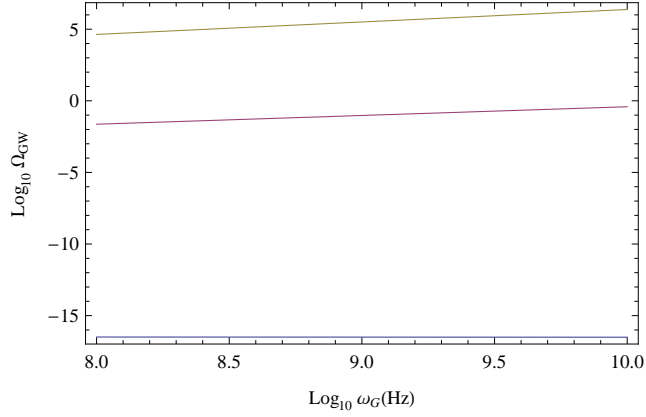}
\caption{The energy density of Relic gravitational waves for various $n_t=-r_t/8,0.61,0.87$ that correspond to the blue curve, the purple curve and the brown curve, respectively.}
\label{fig:density}
\end{figure}

In this letter, to focus on the effect of tensor spectral index $n_t$, we fix the value of tensor-to-scalar ratio $r_t=0.05$ to investigate high frequency RGW. Only the frequency band of RGW that range from $10^8$Hz to $10^{10}$Hz belongs to the detective region of high frequency detector \cite{FYLi}. Thus, we investigate the RGW in this frequency band. In Fig.\ref{fig:amplitude},\ref{fig:density}, we apply the mean value of cosmological parameters \cite{Planck XIII} to give the amplitude $h_t(\eta_0,\omega_G)$ and the numerical results of the energy density $\Omega_{GW}(\omega_G)$, respectively. Here, the frequency $\omega_G$ of RGW and wavenumber $k$ have the following relation $k=2\pi\omega_G$ and the pivot scale is $k_p=0.01 \rm Mpc^{-1}$. We choose $n_t=-r_t/8,0.61,0.87$ to plot Fig.\ref{fig:amplitude},\ref{fig:density} that correspond to the consistency relation of the canonical single-field slow-roll inflation models, 68\% confidence level bound and 95\% confidence level bound that are constrained by combining the data of BKP and LIGO \cite{Huang}, respectively.

\begin{figure}
\includegraphics[scale=1]{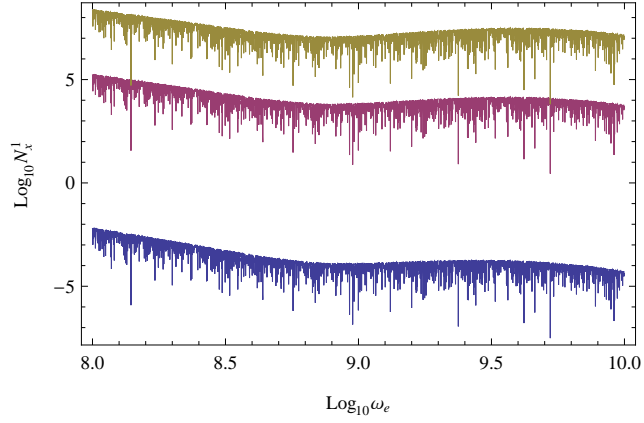}
\caption{The transverse perturbative photon flux $N_x^1\rm(s^{-1})$ at $x=3.5\rm cm$ for various $n_t=-r_t/8,0.61,0.87$ that correspond to the blue curve, the purple curve and the brown curve, respectively.}
\label{fig:Nx1}
\end{figure}

\begin{figure}
\includegraphics[scale=0.7]{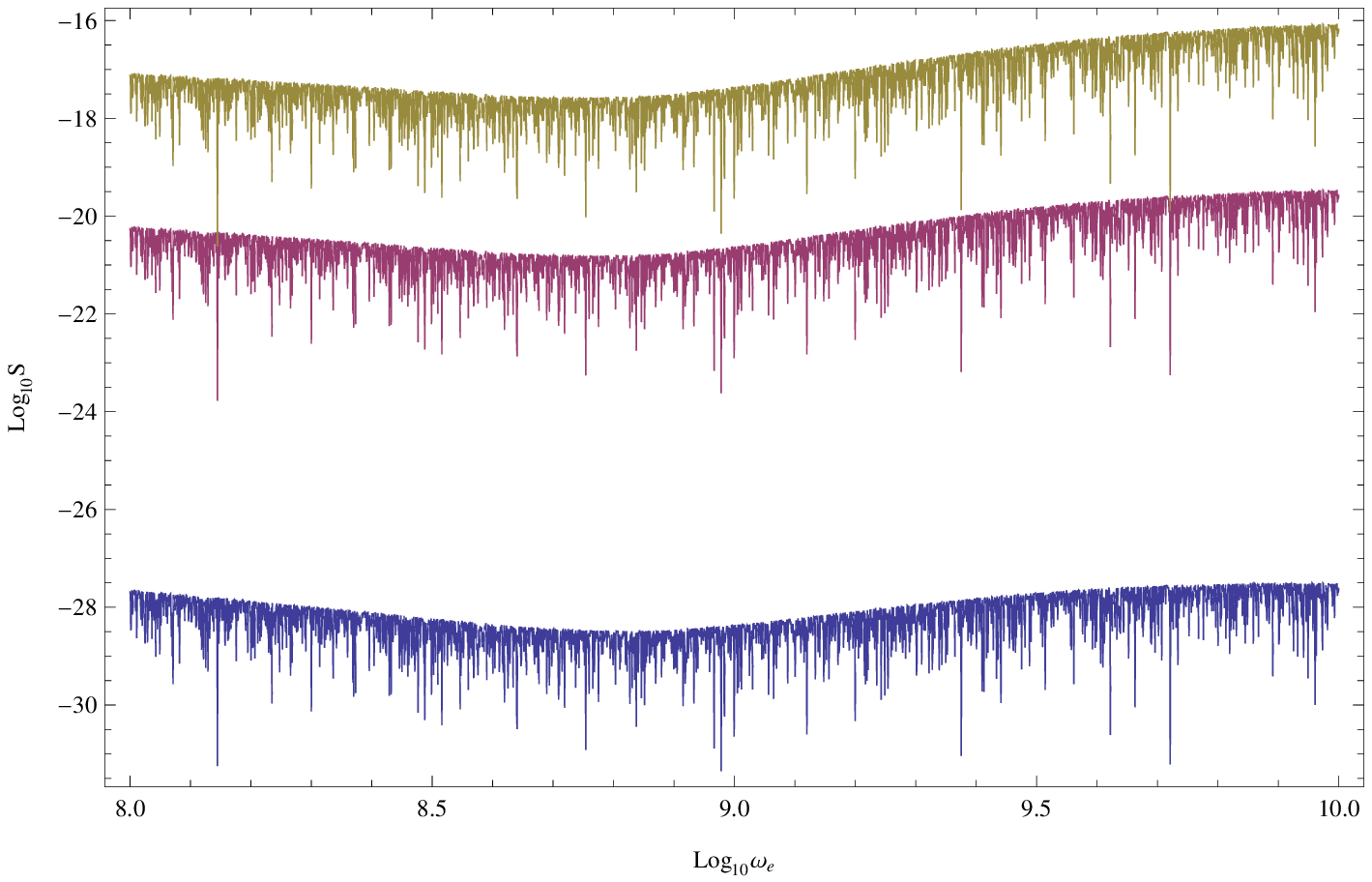}
\caption{The ratio of the transverse perturbative photon flux $N_x^1$ to the background photon flux $N_x^0$, i.e. $S=N_x^1|_{x=3.5\rm cm}/N_x^0|_{x=3.5\rm cm}$, for various $n_t=-r_t/8,0.61,0.87$ that correspond to the blue curve, the purple curve and the brown curve, respectively.}
\label{fig:ratio}
\end{figure}

\section{Expected signals in high frequency GW detector}
F.-Y. Li's high frequency GW detector \cite{FYLi} applies the electromagnetic perturbation effects that produced by high frequency GW to detect GW. The designed experimental device is embedded in a cavity. It is constituted by a Gaussian beam (GB) that passing through a static magnetic field. The power of GB is $10$W, and its frequency mode is given as
\begin{equation}\label{GB}
\psi=\frac{\psi_0\exp(-r^2/W^2)}{\sqrt{1+(z/f)^2}}\exp\left(i\left((k_ez-\omega_et)-\tan^{-1}\frac{z}{f}+\frac{k_er^2}{2R}+\delta\right)\right),
\end{equation}
where $\psi_0\approx 1.26\times 10^3 \rm Vm^{-1}$ denotes the amplitude of GB, $f=k_eW_0^2/2$, $r^2=x^2+y^2$, $W=W_0\sqrt{1+(z/f)^2}$, $W_0=0.05 \rm m^2$ denotes the minimum spot radius and $R=z+f^2/z$ denotes the curvature radius of the wave fronts of the GB at $z$. The static magnetic field $\bar{B}_y^0$ is designed to be $10$T and pointing along the $y$-axis. It is localized in the region $-l_1\leq z\leq l_2$, where $l_1=l_2=0.3\rm m$. The GB (\ref{GB}) does not have electric field that points $z$-axis. For simplicity, we set background electric field $E_x^0=\psi$ and $E_z^0=0$. Thus, from the Maxwell's equation, one can obtain
\begin{eqnarray}
E_y^0&=&2x\left(\frac{1}{W^2}-\frac{ik_e}{2R}\right)\int E_x^0 dy,\\
B_z^0&=&\frac{i}{\omega_e}\left(\frac{\partial E_x^0}{\partial y}-\frac{\partial E_y^0}{\partial x}\right).
\end{eqnarray}
While the high frequency GW that propagate along $z$-axis pass through the cavity, they will produce perturbed electromagnetic field. If the frequency of GW $\omega_G$ equals to the frequency of GB $\omega_e$, the coherent effect magnifies the effect of GB and produces the transverse perturbative photon flux. It has been shown in Ref. \cite{FYLi} that the transverse perturbative photon flux that propagates along $x$-axis is larger than the one that propagates along other directions. In this letter, we only focus on investigating the transverse perturbative photon flux that propagates along $x$-axis.

The dynamics of the electromagnetic system in cavity should satisfy the electromagnetic equations in curved spacetime \cite{Birrel}, since the spacetime is fluctuated by GW. These electromagnetic equations can be solved by perturbed approach for the amplitude of GW is very small. By making use of the solutions of these electromagnetic equations, one can obtain the background photon irradiance and the transverse perturbative photon irradiance. The background photon irradiance along $x$-axis is given as
\begin{equation}\label{nx0}
n_x^0=\frac{1}{\hbar\omega_e}\Big\langle\frac{1}{\mu_0}(E_y^0B_z^0)\Big\rangle,
\end{equation}
where $\mu_0$ is the vacuum permeability.
The transverse perturbative photon irradiance along $x$-axis is given as
\begin{equation}\label{nx1}
n_x^1=\frac{1}{\hbar\omega_e}\Big\langle\frac{1}{\mu_0}(E_y^1B_z^0)\Big\rangle,
\end{equation}
where the perturbed electric field $E_y^1$ that is generated from the GW is given as
\begin{equation}\label{Ey1}
E_y^1=h_t\bar{B}_y^0c\left(-\frac{1}{2}k_e(z+l_1)\exp[i(k_ez-\omega_et)]+\frac{i}{4}\exp[i(k_ez+\omega_et)]\right),
\end{equation}
and $c$ denotes the speed of light. The background photon flux along $x$-axis and the transverse perturbative photon flux along $x$-axis can be derived from the equations (\ref{nx0},\ref{nx1}). They are given as
\begin{eqnarray}
N_x^0&=&\iint_{\Delta s} n_x^0 dy dz,\\
N_x^1&=&\iint_{\Delta s} n_x^1 dy dz,
\end{eqnarray}
where the``typical receiving surface" $\Delta s$ of the high frequency GW detector equals $3\times 10^{-2} \rm m^2$, i.e., $0<y<0.1 \rm m$, $0<z<0.3 \rm m$.

The transverse perturbative photon flux $N_x^1|_{x=3.5\rm cm}$ is shown in Fig.\ref{fig:Nx1}.
The ratio of the transverse perturbative photon flux $N_x^1$ to the background photon flux $N_x^0$, i.e. $S=N_x^1|_{x=3.5\rm cm}/N_x^0|_{x=3.5\rm cm}$, describes the basic feature of high frequency GW detector. It is shown in Fig.~\ref{fig:ratio}. The Fig.~\ref{fig:ratio} implies that the spectrum of background photon flux have different behavior from the spectrum of perturbative photon flux. It means that one can remove the background photon flux away from the observed photon flux.

\section{Conclusions and Remarks}\label{sec:conclusion}
The designed high frequency GW detector as a new window could provide more information for the stage of inflation.
The power spectrum of primordial tensor perturbations $\mathcal{P}_t$ increases rapidly in high frequency region if the spectral index $n_t>0$. Thus, high frequency GW detector could give more severe constraint on tensor spectral index $n_t$. In this letter, we showed the amplitude of RGW with frequency range of $10^8-10^{10}$Hz. It is shown in Fig. \ref{fig:amplitude} that the amplitude of RGW $h_t$($5\times10^9$Hz) varies from $10^{-36}$ to $10^{-25}$ while $n_t$ varies from $-r_t/8$ to $0.87$. Interaction between the RGW and electromagnetic system could generate a perturbed electromagnetic field. F.-Y. Li's high frequency GW detector \cite{FYLi} applies this effect to detect GW. In this letter, we have used the designed experimental parameters to calculate the perturbative photon flux $N_x^1$ that is generated from RGW. We find from Fig. \ref{fig:Nx1} that the perturbative photon flux $N_x^1$ decreases while the frequency increases. And the scale of the cavity is designed to be $60\rm cm$. These facts imply that F.-Y. Li's detector should focus on detecting the RGW with frequency $5\times10^9$Hz such that wavelength of RGW matches the scale of the cavity. The perturbative photon flux $N_x^1|_{x=3.5\rm cm}$($5\times10^9$Hz) varies from $1.40\times10^{-4}\rm s^{-1}$ to $2.85\times10^{7}\rm s^{-1}$ while $n_t$ varies from $-r_t/8$ to $0.87$. In Fig.\ref{fig:ratio}, we have shown the ratio of the transverse perturbative photon flux $N_x^1$ to the background photon flux $N_x^0$ that describes the basic feature of high frequency GW detector. It varies from $10^{-28}$ to $10^{-16}$ while $n_t$ varies from $-r_t/8$ to $0.87$. The Fig. \ref{fig:ratio} implies that the spectrum of background photon flux have different behavior from the spectrum of perturbative photon flux. The signal and background photon fluxes also have other different physical behaviors in specific local regions, such as different flux distribution, propagating direction, phase, wave impedance, decay rate, etc. These distinctive behaviors would provide a potential way to remove the background photon flux away from the observed photon flux.

The noise in high frequency GW detector plays an important role in detecting GW.
In order to detect GW, the ratio of signal to noise should be large enough.
For example, thermal noise can be roughly estimated by $k_B T$ ($k_B$ is Boltzmann constant and $T$ is the temperature of thermal noise). Then, $T<\hbar\omega_G/k_b\sim0.24K$ if $\omega_G=5\times10^9$Hz.
However, the noise strongly depends on the experimental setup.
Thus, the detail analysis of the noise in high frequency GW detector shall be carried out in future work.

\vspace{1cm}
\begin{acknowledgments}
We thank prof. F.-Y. Li for useful discussions. X.Li has been supported by the National Natural Science Fund of China (NSFC) (Grant NO. 11305181) and the Open Project Program of State Key Laboratory of Theoretical Physics, Institute of Theoretical Physics, Chinese Academy of Sciences, China (No. Y5KF181CJ1). S.Wang has been supported by grants from NSFC (Grant NO. 11322545 and 11335012). H. Wen has been supported by the Fundamental Research Fund for the Central Universities (Project No. 106112015CDJRC131216).
\end{acknowledgments}

\end{document}